\documentclass[preprint]{ptephy}

\preprintnumber{KYUSHU-HET-182}

\usepackage{amssymb}
\usepackage{amsthm}
\usepackage{amsmath}
\usepackage{booktabs}
\usepackage{bbm}
\usepackage{mathtools}
\DeclareMathOperator{\tr}{tr}

\newcommand{\Slash}[1]{{\ooalign{\hfil/\hfil\crcr$#1$}}}
\numberwithin{equation}{section}

\begin{document}

\title{Gradient flow and the Wilsonian renormalization group flow}

\author{%
\name{\fname{Hiroki} \surname{Makino}}{1},
\name{\fname{Okuto} \surname{Morikawa}}{1},
and
\name{\fname{Hiroshi} \surname{Suzuki}}{1,\ast}
}

\address{%
\affil{1}{Department of Physics, Kyushu University
744 Motooka, Nishi-ku, Fukuoka, 819-0395, Japan}
\email{hsuzuki@phys.kyushu-u.ac.jp}
}

\begin{abstract}
The gradient flow is the evolution of fields and physical quantities along a
dimensionful parameter~$t$, the flow time. We give a simple argument that
relates this gradient flow and the Wilsonian renormalization group (RG) flow.
We then illustrate the Wilsonian RG flow on the basis of the gradient flow in
two examples that possess an infrared fixed point, the 4D many-flavor gauge
theory and the 3D $O(N)$ linear sigma model.
\end{abstract}

\subjectindex{B32, B01, B37}
\maketitle

\section{Introduction and the basic idea}
\label{sec:1}
The gradient flow~\cite{Narayanan:2006rf,Luscher:2009eq,Luscher:2010iy,%
Luscher:2011bx,Luscher:2013cpa} is the evolution of fields and physical
quantities along a dimensionful parameter~$t$, the flow time; the flow acts as
the ``coarse-graining'' as $t>0$ becomes large. These two features of the
gradient flow are common to the Wilsonian renormalization group (RG)
flow~\cite{Wilson:1973jj} in a broad sense, provided that the flow time is
identified with the renormalization scale. In fact, it has sometimes been
indicated that the gradient flow and the Wilsonian RG flow can be identified in
some
ways~\cite{Luscher:2013vga,Kagimura:2015via,Yamamura:2015kva,Aoki:2016ohw};
see also Refs.~\cite{Aoki:2015dla,Aoki:2016env,Aoki:2017bru,Aoki:2017uce} for
related studies. In this paper, we give a simple argument that relates the
gradient flow and the Wilsonian RG flow; our argument is somewhat similar to
that of~Ref.~\cite{Luscher:2013vga}. We then illustrate the Wilsonian RG flow
on the basis of the gradient flow in two examples that possess an infrared
fixed point, the 4D many-flavor gauge theory and the 3D $O(N)$ linear sigma
model.

Our idea is very simple. We take the following flow equations for the gauge
potential~$A_\mu(x)$ and for the Dirac
fields $\psi(x)$ and~$\Bar{\psi}(x)$:\footnote{Here, the covariant derivative
on the gauge field are defined by $D_\mu\equiv\partial_\mu+[B_\mu,\cdot]$; the
field strength is defined by
$G_{\mu\nu}(t,x)\equiv\partial_\mu B_\nu(t,x)-\partial_\nu B_\mu(t,x)%
+[B_\mu(t,x),B_\nu(t,x)]$. The Laplacians on the Dirac fields is defined by
$\Delta\equiv D_\mu D_\mu$, and
$\overleftarrow{\Delta}\equiv\overleftarrow{D}_\mu\overleftarrow{D}_\mu$
from the covariant derivatives on the Dirac fields,
$D_\mu=\partial_\mu+B_\mu$
and~$\overleftarrow{D}_\mu\equiv\overleftarrow{\partial}_\mu-B_\mu$.
We will occasionally use notation such as $A_\mu(x)=A_\mu^a(x)T^a$ by using the
generator of the gauge group, $T^a$.}
\begin{align}
   \partial_tB_\mu(t,x)&=D_\nu G_{\nu\mu}(t,x),&
   B_\mu(t=0,x)&=A_\mu(x),
\label{eq:(1.1)}
\\
   \partial_t\chi(t,x)&=\Delta\chi(t,x),&
   \chi(t=0,x)&=\psi(x),
\label{eq:(1.2)}
\\
   \partial_t\Bar{\chi}(t,x)
   &=\Bar{\chi}(t,x)
   \overleftarrow{\Delta},&
   \Bar{\chi}(t=0,x)&=\Bar{\psi}(x).
\label{eq:(1.3)}
\end{align}
Let us consider the correlation function of operators composed of the flowed
fields:
\begin{equation}
   \left\langle\mathcal{O}_1(t_1,x_1)\dotsb\mathcal{O}_N(t_N,x_N)
   \right\rangle.
\label{eq:(1.4)}
\end{equation}
Let us also suppose that we have a set of (a generally infinite number of)
coupling constants~$\{g_i\}$ with which the correlation function is
computed.\footnote{We implicitly assume the presence of the ultraviolet cutoff.}
We consider the mapping in this space of the coupling constants induced by
the Wilsonian RG flow,
\begin{equation}
   \{g_i\}\to\{g_i(\xi)\},
\label{eq:(1.5)}
\end{equation}
where $\xi$ parametrizes the RG flow. This RG flow can be characterized by the
scaling relation\footnote{Here, we neglect a possible non-trivial mixing of
operators under the RG flow, for notational simplicity.}
\begin{equation}
   \left\langle\mathcal{O}_1(e^{2\xi}t_1,e^\xi x_1)\dotsb
   \mathcal{O}_N(e^{2\xi}t_N,e^\xi x_N)
   \right\rangle_{\{g_i\}}
   =Z(\xi)\left\langle\mathcal{O}_1(t_1,x_1)\dotsb
   \mathcal{O}_N(t_N,x_N)
   \right\rangle_{\{g_i(\xi)\}},
\label{eq:(1.6)}
\end{equation}
where $Z(\xi)$ is the multiplicative renormalization factor and the
subscript implies that the correlation function is evaluated with respect to
the set of coupling constants. Compare this relation with, for instance,
Eqs.~(7.10) and~(7.15) of~Ref.~\cite{Wilson:1973jj}. Note that the flow time
has the mass dimension~$-2$ instead of~$-1$. The advantage of this
characterization of the Wilsonian RG flow is that this scaling relation itself
can be written down even for gauge theory for which the momentum cutoff is
incompatible with the gauge invariance (at least naively). In particular, for
the one-point function of an operator that does not require the multiplicative
renormalization, 
\begin{equation}
   \left\langle\mathcal{O}_1(e^{2\xi}t)\right\rangle_{\{g_i\}}
   =\left\langle\mathcal{O}_1(t)\right\rangle_{\{g_i(\xi)\}},
\label{eq:(1.7)}
\end{equation}
where we have omitted the argument~$x$ assuming translational invariance in
the $x$-space. Hence, assuming that the correspondence,
\begin{equation}
   \left\{\left\langle\mathcal{O}_i(t)\right\rangle\right\}
   \Leftrightarrow\{g_i(\xi)\}
\label{eq:(1.8)}
\end{equation}
arising from~Eq.~\eqref{eq:(1.7)} is one to one, we can use the one-point
functions $\{\langle\mathcal{O}_i(t)\rangle\}$ instead of the coupling
constants~$\{g_i(\xi)\}$. Of course, this idea is well known for the case of
the gauge coupling constant~\cite{Luscher:2010iy}:
\begin{equation}
   g^2(\mu=1/\sqrt{8t})\propto t^2\left\langle
   G_{\mu\nu}^aG_{\mu\nu}^a(t)\right\rangle.
\label{eq:(1.9)}
\end{equation}

In what follows, we illustrate the idea~\eqref{eq:(1.8)} in theories in which
several coupling constants play an interesting role; we will observe the flow
of relevant and irrelevant coupling constants around an RG fixed point through
the correspondence~\eqref{eq:(1.8)}. We hope that our present consideration
will be useful for more difficult models for which an infrared non-trivial
fixed point can be concluded only non-perturbatively.

\section{4D $N_f$-flavor gauge theory and the Banks--Zaks fixed point}
\label{sec:2}
Our first example is the 4D vector-like gauge theory with $N_f$-flavor Dirac
fermions with the degenerate mass~$m$. As the operators
in~Eq.~\eqref{eq:(1.8)}, we take (as the one corresponding to the gauge
coupling~\cite{Luscher:2010iy})%
\footnote{The generators~$T^a$ ($a$ runs from~$1$ to~$\dim(G)$) of the gauge
group~$G$ are anti-Hermitian and the structure constants are defined
by~$[T^a,T^b]=f^{abc}T^c$. Quadratic Casimirs are defined
by~$f^{acd}f^{bcd}=C_2(G)\delta^{ab}$ and, for a gauge representation~$R$,
$\tr_R(T^aT^b)=-T(R)\delta^{ab}$ and~$T^aT^a=-C_2(R)1$. We also denote
$\tr_R(1)=\dim(R)$.}
\begin{equation}
   \mathcal{O}_1(t,x)
   \equiv\frac{8(4\pi)^2t^2}{3\dim(G)}
   \frac{1}{4}G_{\mu\nu}^a(t,x)G_{\mu\nu}^a(t,x),
\label{eq:(2.1)}
\end{equation}
and
\begin{equation}
   \mathcal{O}_2(t,x)
   \equiv\frac{\Bar{\chi}(t,x)\chi(t,x)}
   {t^{1/2}\left\langle\Bar{\chi}(t,x)\overleftrightarrow{\Slash{D}}
   \chi(t,x)\right\rangle}
   \equiv-\frac{(4\pi)^2t^{3/2}}{2\dim(R)N_f}
   \mathring{\Bar{\chi}}(t,x)\mathring{\chi}(t,x).
\label{eq:(2.2)}
\end{equation}
The flowed gauge field and its local products such as $\mathcal{O}_1(t,x)$ do
not receive any multiplicative renormalization~\cite{Luscher:2011bx}. On
the other hand, although the flowed Dirac field is multiplicatively
renormalized~\cite{Luscher:2013cpa}, the renormalization of local products
is simply determined by the number of Dirac fields in the product. Thus,
$\mathcal{O}_2(t,x)$ in~Eq.~\eqref{eq:(2.2)} also does not receive
multiplicative renormalization because of the division by the expectation
value~\cite{Makino:2014taa}.

In the present system, one observes the so-called Banks--Zaks infrared fixed
point~\cite{Caswell:1974gg,Banks:1981nn} if one uses the two-loop approximation
of the beta function. We introduce the running gauge coupling~$\Bar{g}(\mu)$
and the running mass parameter~$\Bar{m}(\mu)$ in the $\overline{\text{MS}}$
scheme, respectively, by
\begin{align}
   &\left[b_0\Bar{g}(\mu)^2\right]^{-b_1/(2b_0^2)}
   \exp\left[-\frac{1}{2b_0\Bar{g}(\mu)^2}\right]=\frac{\Lambda}{\mu},
\label{eq:(2.3)}
\\
   &\Bar{m}(\mu)=M\left[2b_0\Bar{g}(\mu)^2\right]^{d_0/(2b_0)},
\label{eq:(2.4)}
\end{align}
where
\begin{align}
   b_0&=\frac{1}{(4\pi)^2}
   \left[\frac{11}{3}C_2(G)-\frac{4}{3}T(R)N_f\right],
\label{eq:(2.5)}
\\
   b_1&=\frac{1}{(4\pi)^4}
   \left\{
   \frac{34}{3}C_2(G)^2-\left[4C_2(R)+\frac{20}{3}C_2(G)\right]T(R)N_f
   \right\},
\label{eq:(2.6)}
\\
   d_0&=\frac{1}{(4\pi)^2}6C_2(R),
\label{eq:(2.7)}
\end{align}
and $\Lambda$ and~$M$ are RG invariant mass scales. In terms of these running
parameters, we have the one-point function,
\begin{equation}
   \left\langle\mathcal{O}_1(t)\right\rangle
   =\Bar{g}(1/\sqrt{8t})^2
   \left[1+\frac{\Bar{g}(1/\sqrt{8t})^2}{(4\pi)^2}K_1(t)
   +\frac{\Bar{g}(1/\sqrt{8t})^4}{(4\pi)^4}K_2
   \right],
\label{eq:(2.8)}
\end{equation}
where
\begin{equation}
   K_1(t)=
   \left(\frac{11}{3}\gamma_E+\frac{52}{9}-3\ln3\right)C_2(G)
   +\left[-\frac{4}{3}\gamma_E-\frac{8}{9}+\frac{8}{3}\ln2
   +16\Bar{m}(1/\sqrt{8t})^2t
   \right]T(R)N_f,
\label{eq:(2.9)}
\end{equation}
and
\begin{align}
   K_2&=8(4\pi)^2
   \bigl\{
   -0.0136423(7)C_2(G)^2
\notag\\
   &\qquad\qquad\qquad{}
   +\left[
   0.006\,440\,134(5)C_2(R)-0.008\,688\,4(2)C_2(G)
   \right]T(R)N_f
\notag\\
   &\qquad\qquad\qquad\qquad{}
   +0.000\,936\,117T(R)^2N_f^2
   \bigr\}.
\label{eq:(2.10)}
\end{align}
Equation~\eqref{eq:(2.8)} for the massless case was obtained
in~Ref.~\cite{Luscher:2010iy} and for general mass cases
in~Ref.~\cite{Harlander:2016vzb}; we have retained only the leading mass
correction in~Eq.~\eqref{eq:(2.9)} (as given in~Eq.~(2.34)
of~Ref.~\cite{Harlander:2016vzb}). Although this treatment of the mass
correction, which is also adopted in~Eq.~\eqref{eq:(2.11)}, is approximate,
this makes the resulting RG equations~\eqref{eq:(2.12)} and~\eqref{eq:(2.13)}
quite simple and illustrative, so here we content ourselves with this
approximate treatment.

On the other hand, to the one-loop order, $\langle\mathcal{O}_2(t,x)\rangle$
is given by\footnote{The computation of this is given in~v2 of the arXiv
reference in~Ref.~\cite{Makino:2014taa}}
\begin{equation}
   \left\langle\mathcal{O}_2(t)\right\rangle
   =\Bar{m}(1/\sqrt{8t})t^{1/2}
   \left\{1+\frac{\Bar{g}(1/\sqrt{8t})^2}{(4\pi)^2}
   \left[3\gamma_E+4+2\ln2-\ln(432)\right]C_2(R)
   \right\}.
\label{eq:(2.11)}
\end{equation}

We now take the flow time derivatives of Eqs.~\eqref{eq:(2.8)}
and~\eqref{eq:(2.11)}. By using Eqs.~\eqref{eq:(2.3)} and~\eqref{eq:(2.4)} (or
the corresponding RG equations) and eliminating the running parameters in favor
of one-point functions, we arrive at
\begin{align}
   t\frac{d}{dt}\left\langle\mathcal{O}_1(t)\right\rangle
   &=b_0\left\langle\mathcal{O}_1(t)\right\rangle^2
   +b_1\left\langle\mathcal{O}_1(t)\right\rangle^3
   +\frac{1}{(4\pi)^2}16T(R)N_f
   \left\langle\mathcal{O}_1(t)\right\rangle^2
   \left\langle\mathcal{O}_2(t)\right\rangle^2,
\label{eq:(2.12)}
\\
   t\frac{d}{dt}\left\langle\mathcal{O}_2(t)\right\rangle
   &=\frac{1}{2}
   \left[1+d_0\left\langle\mathcal{O}_1(t)\right\rangle\right]
   \left\langle\mathcal{O}_2(t)\right\rangle.
\label{eq:(2.13)}
\end{align}
From these equations, it is clear that $\langle\mathcal{O}_1(t)\rangle$
and $\langle\mathcal{O}_2(t)\rangle$ can be used as parameters in the
coupling constant space. Note that the RG coefficients $b_0$, $b_1$ and~$d_0$
are universal. In the infrared limit $t\to\infty$,
$\langle\mathcal{O}_2(t)\rangle\neq0$ corresponds to a relevant coupling
$\langle\mathcal{O}_2(t)\rangle\to\infty$ around the Banks-Zaks fixed point
at $(\langle\mathcal{O}_1(t)\rangle,\langle\mathcal{O}_2(t)\rangle)=%
(-b_0/b_1,0)$.\footnote{We assume $b_1<0$; note that
$\langle\mathcal{O}_1(t)\rangle>0$ by definition.}

\section{3D $O(N)$ linear sigma model at large~$N$ and the Wilson--Fisher
fixed point}
\label{sec:3}
Our second example is the 3D $O(N)$ linear sigma model that possesses the
so-called Wilson--Fisher fixed point~\cite{Wilson:1971dc} in the infrared
limit. The gradient flow of an operator in this system in relation to the
Wilsonian RG flow was studied in detail in~Ref.~\cite{Aoki:2016ohw} and the
Wilson--Fisher fixed point was observed. Actually, our present study was
partially motivated by the study of~Ref.~\cite{Aoki:2016ohw}. We will consider
the RG flow in the 2D coupling constant space in which there is one direction
of the relevant operator around the fixed point (in~Ref.~\cite{Aoki:2016ohw},
only 1D space along the irrelevant coupling is considered). We will work out
the large-$N$ approximation to the order of our concern. So, we first
recapitulate the solution of the model in the large-$N$ approximation for later
use.

\subsection{The solution in the large-$N$ approximation}
\label{sec:3.1}
The Euclidean action of the 3D $O(N)$ linear sigma model is given by
\begin{equation}
   S=\int d^3x\,
   \left\{
   \frac{1}{2}\partial_\mu\phi_i(x)\partial_\mu\phi_i(x)
  +\frac{1}{2}m_0^2\phi_i(x)\phi_i(x)
   +\frac{1}{8N}\lambda_0\left[\phi_i(x)\phi_i(x)\right]^2
   \right\},
\label{eq:(3.1)}
\end{equation}
where $i=1$, \dots, $N$. We introduce the effective action, i.e., the
generating functional of the 1PI correlation functions, as
\begin{equation}
   {\mit\Gamma}[\phi]
   =\sum_{n=0}^\infty\frac{1}{n!}\int d^3x_1\dotsb d^3x_n\,
   \phi_{i_1}(x_1)\dotsb\phi_{i_n}(x_n)
   {\mit\Gamma}_{i_1\dotsb i_n}^{(n)}(x_1,\dotsc,x_n),
\label{eq:(3.2)}
\end{equation}
where ${\mit\Gamma}_{i_1\dotsb i_n}^{(n)}(x_1,\dotsc,x_n)$ are the vertex
functions. We also introduce the Fourier transformation:
\begin{align}
   &{\mit\Gamma}_{i_1\dotsb i_n}^{(n)}(x_1,\dotsc,x_n)
\notag\\
   &=\int\frac{d^3p_1}{(2\pi)^3}\dotsb\frac{d^3p_n}{(2\pi)^3}\,
   e^{-ip_1x_1-\dotsb-ip_nx_n}
   \Tilde{{\mit\Gamma}}_{i_1\dotsb i_n}^{(n)}(p_1,\dotsc,p_n)
   (2\pi)^3\delta(p_1+\dotsb+p_n).
\label{eq:(3.3)}
\end{align}

The large-$N$ approximation in this model is well known and, at the leading
order of the approximation, by using the auxiliary field method for instance,
we have
\begin{align}
   &\Tilde{{\mit\Gamma}}_{i_1i_2}^{(2)}(p_1,p_2)
   =\delta_{i_1i_2}(p_1^2+M^2),
\label{eq:(3.4)}
\\
   &\Tilde{{\mit\Gamma}}_{i_1i_2i_3i_4}^{(4)}(p_1,p_2,p_3,p_4)
\notag\\
   &=\delta_{i_1i_2}\delta_{i_3i_4}\frac{\lambda_0}{N}
   \left[1
   +\frac{\lambda_0}{8\pi}
   \frac{1}{\sqrt{(p_1+p_2)^2}}
   \arctan\left(\frac{1}{2}\sqrt{\frac{(p_1+p_2)^2}{M^2}}\right)
   \right]^{-1}
   +(2\leftrightarrow3)+(2\leftrightarrow4).
\label{eq:(3.5)}
\end{align}
In these expressions, the ``physical'' mass~$M$ is given by the solution to the
so-called gap equation,
\begin{equation}
   M^2+\frac{\lambda_0}{8\pi}M=m_0^2+\frac{1}{4\pi^2}\lambda_0\Lambda,
\label{eq:(3.6)}
\end{equation}
with $\Lambda$ being the momentum cutoff.

In the present model, the renormalized parameters in the mass-independent
renormalization scheme can be defined as
\begin{equation}
   m_0^2=Z_m m^2+\delta m_0^2,\qquad \lambda_0=Z_\lambda\lambda.
\label{eq:(3.7)}
\end{equation}
As Eq.~\eqref{eq:(3.4)} shows, there is no need of the wave function
renormalization in the leading order of the large-$N$ approximation. We fix the
renormalization constants $Z_m$, $\delta m_0^2$, and~$Z_\lambda$ by imposing the
following renormalization conditions at the renormalization scale~$\mu$:
\begin{align}
   &\left.\Tilde{{\mit\Gamma}}_{i_1i_2}^{(2)}(p_1,p_2)\right|_{p_1^2=p_2^2=0,m^2=0}
   =0,
\label{eq:(3.8)}
\\
   &\left.\Tilde{{\mit\Gamma}}_{i_1i_2}^{(2)}(p_1,p_2)
   \right|_{p_1^2=p_2^2=0,m^2=\mu^2}
   =\mu^2,
\label{eq:(3.9)}
\\   
   &\left.\Tilde{{\mit\Gamma}}_{i_1i_2i_3i_4}^{(4)}
   (p_1,p_2,p_3,p_4)
   \right|_{p_i\cdot p_j=\mu^2\delta_{ij}-\frac{1}{3}\mu^2(1-\delta_{ij}),m^2=\mu^2}
   =\delta_{i_1i_2}\delta_{i_3i_4}\frac{\lambda}{N}
   +(2\leftrightarrow3)+(2\leftrightarrow4).
\label{eq:(3.10)}
\end{align}
From the first two relations, we have
\begin{equation}
   \delta m_0^2=-\frac{1}{4\pi^2}\lambda_0\Lambda,\qquad
   Z_m=1+\frac{1}{8\pi}\frac{\lambda_0}{\mu},
\label{eq:(3.11)}
\end{equation}
and from the last renormalization condition,
\begin{equation}
   \frac{\lambda}{\mu}
   =\frac{\lambda_0}{\mu}
   \left(1+\frac{\sqrt{3}}{96}\frac{\lambda_0}{\mu}\right)^{-1}.
\label{eq:(3.12)}
\end{equation}
This gives rise to the beta function\footnote{The subscript~$0$
in~$(\mu\frac{\partial}{\partial\mu})_0$ implies that the derivative is taken
while the bare parameters are kept fixed.}
\begin{equation}
   \beta\left(\frac{\lambda}{\mu}\right)
   \equiv\left(\mu\frac{\partial}{\partial\mu}\right)_0\frac{\lambda}{\mu}
   =-\frac{\lambda}{\mu}
   +\frac{\sqrt{3}}{96}\left(\frac{\lambda}{\mu}\right)^2.
\label{eq:(3.13)}
\end{equation}
We note that the slopes of the beta function at two zeros of the beta function
(fixed points) are given by
\begin{equation}
   \beta'\left(\frac{\lambda_*}{\mu}=0\right)=-1,\qquad
   \beta'\left(\frac{\lambda_*}{\mu}=\frac{96}{\sqrt{3}}\right)=+1,
\label{eq:(3.14)}
\end{equation}
respectively.

On the other hand, from the above relations, we have
\begin{equation}
   \frac{m^2}{\mu^2}
   =\left(1+\frac{1}{8\pi}\frac{\lambda_0}{\mu}\right)^{-1}
   \left(\frac{m_0^2}{\mu^2}
   +\frac{1}{4\pi^2}\frac{\lambda_0\Lambda}{\mu^2}\right),
\label{eq:(3.15)}
\end{equation}
and
\begin{equation}
   \left(\mu\frac{\partial}{\partial\mu}\right)_0\frac{m^2}{\mu^2}
   =-2\left[
   \frac{1+\left(\frac{3}{2}\frac{1}{8\pi}-\frac{\sqrt{3}}{96}\right)
   \frac{\lambda}{\mu}}
   {1+\left(\frac{1}{8\pi}-\frac{\sqrt{3}}{96}\right)
   \frac{\lambda}{\mu}}
   \right]\frac{m^2}{\mu^2}.
\label{eq:(3.16)}
\end{equation}
This RG equation becomes quite simple in terms of the parameter~$M$ defined
by~Eq.~\eqref{eq:(3.6)}:
\begin{equation}
   \left(\mu\frac{\partial}{\partial\mu}\right)_0\frac{M}{\mu}
   =-\frac{M}{\mu}.
\label{eq:(3.17)}
\end{equation}

\subsection{The flowed system and the RG flow}
\label{sec:3.2}
We now examine the picture~\eqref{eq:(1.8)} in the present model. We first
have to introduce the flow equation for the scalar field~$\phi_i(x)$. The
simplest choice is 
\begin{equation}
   \partial_t\varphi_i(t,x)=\partial_\mu\partial_\mu\varphi_i(t,x),\qquad
   \varphi_i(t=0,x)=\phi_i(x).
\label{eq:(3.18)}
\end{equation}
We refer the reader to~Ref.~\cite{Capponi:2015ucc} for the renormalizability
of the flowed scalar theory. With the above choice, the correlation functions
of the flowed field~$\varphi_i(t,x)$ can be obtained from those of~$\phi_i(y)$
simply substituting $\varphi_i(t,x)$ by
\begin{equation}
   \varphi_i(t,x)=\int d^3y\,
   \int\frac{d^3p}{(2\pi)^3}\,e^{ip(x-y)}e^{-tp^2}\phi_i(y).
\label{eq:(3.19)}
\end{equation}

We thus have, for instance,
\begin{align}
   \left\langle\varphi_i(t,x)\varphi_i(t,x)\right\rangle
   &=Nt^{-1/2}\int\frac{d^3p}{(2\pi)^3}\,\frac{e^{-2p^2}}{p^2+M^2t}
   +O((1/N)^0)
\notag\\
   &\stackrel{t\to0}{\to}N\frac{1}{2(2\pi)^{3/2}}t^{-1/2}-N\frac{M}{4\pi}+O((1/N)^0),
\label{eq:(3.20)}
\\
   \left\langle\partial_\mu\varphi_i(t,x)\partial_\mu\varphi_i(t,x)
   \right\rangle
   &=N\frac{1}{(8\pi)^{3/2}}t^{-3/2}
   -M^2\left\langle\varphi_i(t,x)\varphi_i(t,x)\right\rangle
   +O((1/N)^0),
\label{eq:(3.21)}
\end{align}
and
\begin{align}
   &\left\langle\left[\varphi_i(t,x)\varphi_i(t,x)\right]^2\right\rangle
   -\left(1+\frac{2}{N}\right)
  \left\langle\varphi_i(t,x)\varphi_i(t,x)\right\rangle^2
\notag\\
   &=-N\lambda_0t^{-1/2}\prod_{i=1}^4
   \left(\int\frac{d^3p_i}{(2\pi)^3}\,\frac{e^{-p_i^2}}{p_i^2+M^2t}\right)
   (2\pi)^3\delta(p_1+p_2+p_3+p_4)
\notag\\
   &\qquad{}
   \times
   \left[
   1+\frac{\lambda_0t^{1/2}}{8\pi}\frac{1}{\sqrt{(p_1+p_2)^2}}
   \arctan\left(\frac{1}{2}\sqrt{\frac{(p_1+p_2)^2}{M^2t}}\right)
   \right]^{-1}+O((1/N)^0).
\label{eq:(3.22)}
\end{align}
Note that in these expressions, momentum variables are dimensionless.

\begin{figure}[ht]
\begin{center}
\includegraphics[width=10cm,clip]{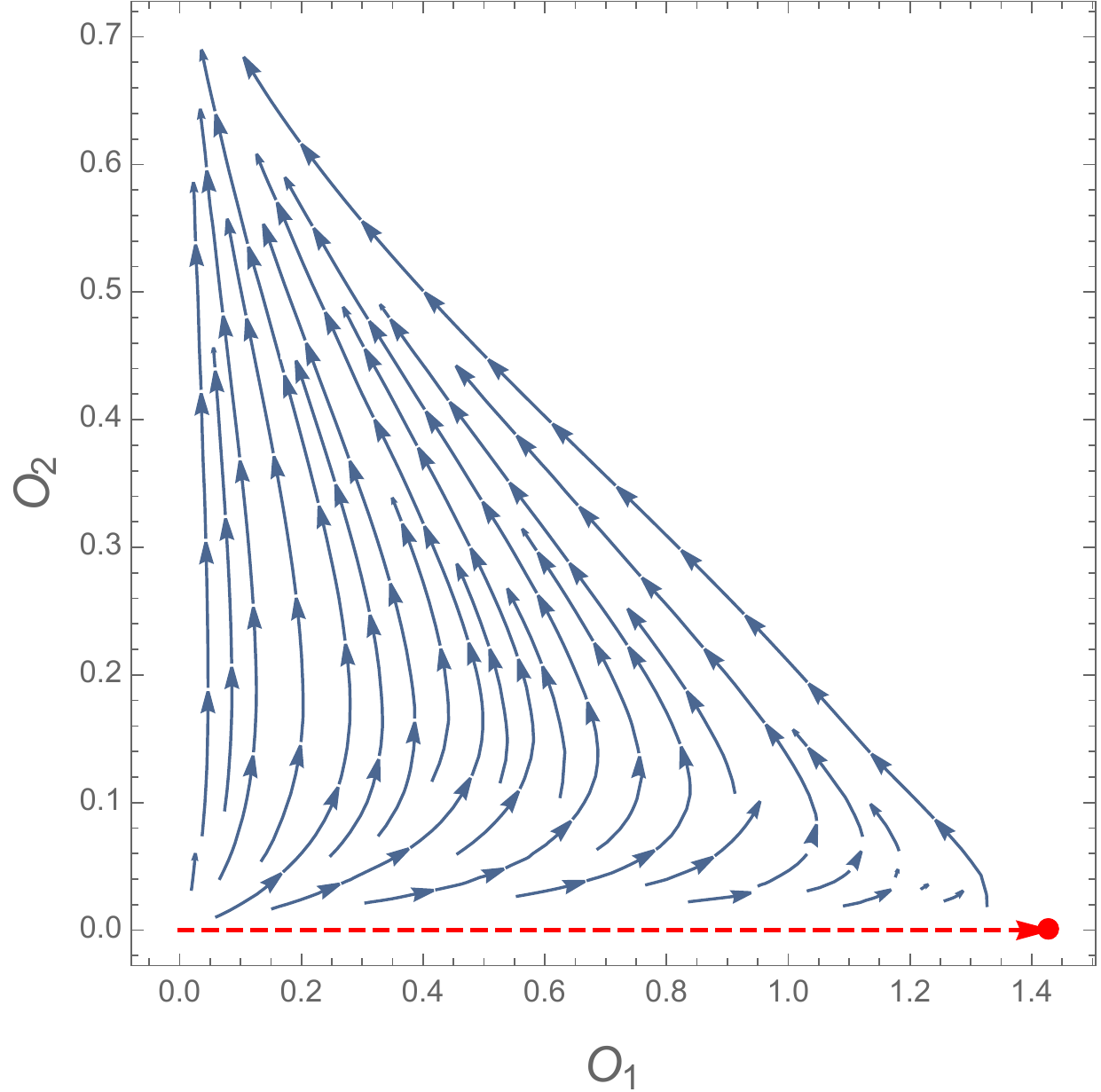}
\caption{The RG flow in the space of $\langle\mathcal{O}_1(t)\rangle$
and~$\langle\mathcal{O}_2(t)\rangle$. The arrows indicate how the point
$(\langle\mathcal{O}_1(t)\rangle,\langle\mathcal{O}_2(t)\rangle)$ changes
as $t$ increases. The red point is the infrared fixed point.}
\label{fig:1}
\end{center}
\end{figure}

\begin{figure}[ht]
\begin{center}
\includegraphics[width=10cm,clip]{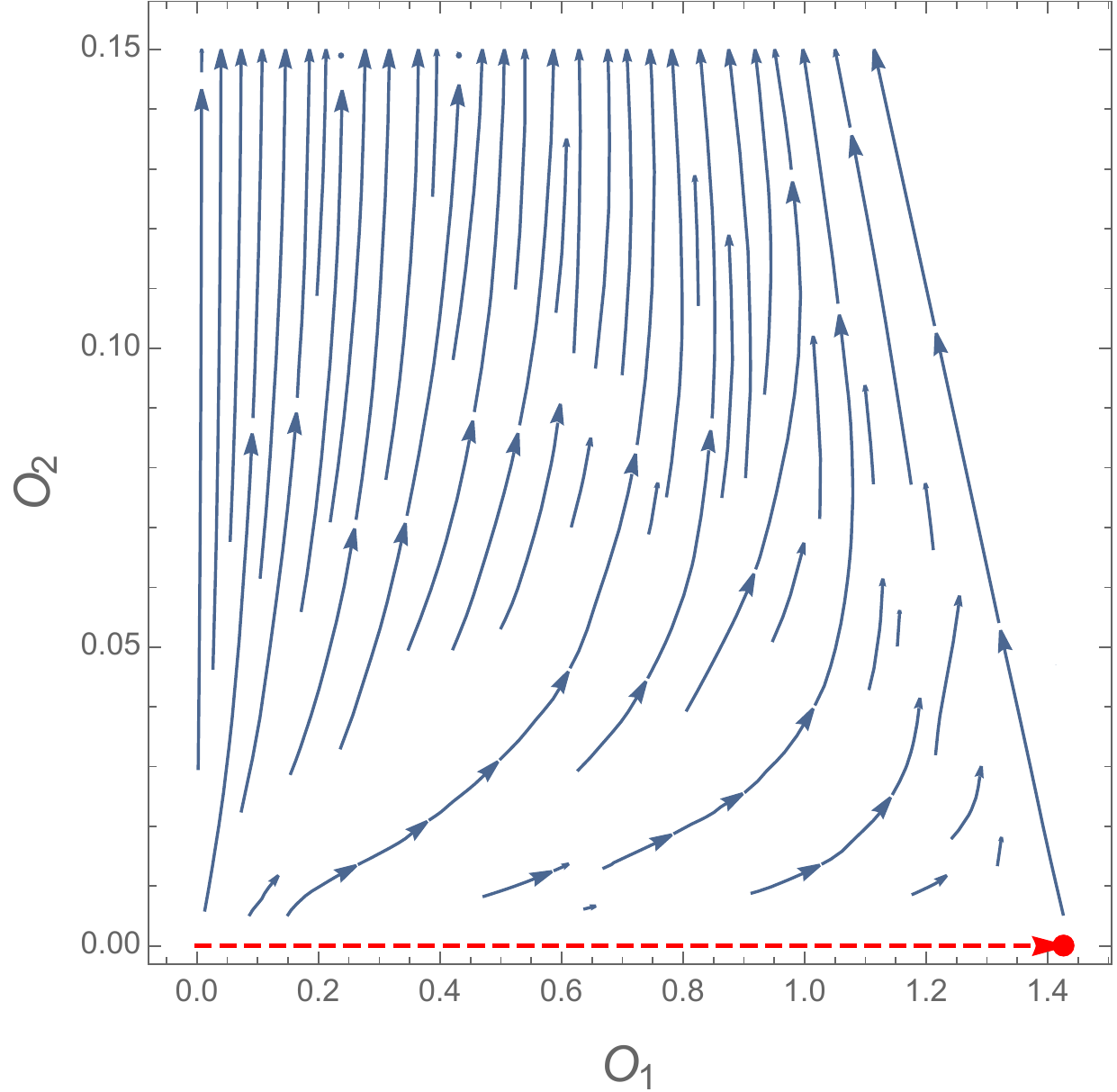}
\caption{Same as Fig.~\ref{fig:1}, but the region around the horizontal axis is
magnified.}
\label{fig:2}
\end{center}
\end{figure}

It is convenient to introduce a new field variable,
\begin{equation}
   \mathring{\varphi}_i(t,x)
   \equiv\sqrt{
   \frac{N}
   {2(2\pi)^{3/2}t^{1/2}\left\langle\varphi_j(t,x)\varphi_j(t,x)\right\rangle}}
   \,\varphi_i(t,x)
   \stackrel{t\to0}{\to}\varphi_i(t,x)+O(1/N),
\label{eq:(3.23)}
\end{equation}
by analogy with~Eq.~\eqref{eq:(2.2)}, which is free from the wave function
renormalization. Using this new variable, we define dimensionless operators,
\begin{align}
   \mathcal{O}_1(t,x)&\equiv
   -\frac{4(2\pi)^3}{N}t
   \left[\mathring{\varphi}_i(t,x)\mathring{\varphi}_i(t,x)\right]^2+(N+2),
\label{eq:(3.24)}
\\
   \mathcal{O}_2(t,x)&\equiv
   \frac{16\pi}{N}t^{3/2}\partial_\mu\mathring{\varphi}_i(t,x)
   \partial_\mu\mathring{\varphi}_i(t,x)
   -\frac{1}{(2\pi)^{1/2}}.
\label{eq:(3.25)}
\end{align}
Then, we have
\begin{align}
   \left\langle\mathcal{O}_1(t)\right\rangle
   &=\lambda_0t^{1/2}
   \left[\int\frac{d^3p}{(2\pi)^3}\,\frac{e^{-2p^2}}{p^2+M^2t}\right]^{-2}
\notag\\
   &\qquad\times\prod_{i=1}^4
   \left(\int\frac{d^3p_i}{(2\pi)^3}\,\frac{e^{-p_i^2}}{p_i^2+M^2t}\right)
   (2\pi)^3\delta(p_1+p_2+p_3+p_4)
\notag\\
   &\qquad\qquad{}
   \times
   \left[
   1+\frac{\lambda_0t^{1/2}}{8\pi}\frac{1}{\sqrt{(p_1+p_2)^2}}
   \arctan\left(\frac{1}{2}\sqrt{\frac{(p_1+p_2)^2}{M^2t}}\right)
   \right]^{-1}
\label{eq:(3.26)}
\\
   &\stackrel{t\to0}{\to}K\lambda_0t^{1/2}
\label{eq:(3.27)}
\\
   &\stackrel{t\to\infty}{\to}
   K'\frac{\lambda_0}{M}\left(
   1+\frac{1}{16\pi }\frac{\lambda_0}{M}
   \right)^{-1}
   \frac{1}{M^3t^{3/2}},
\label{eq:(3.28)}
\end{align}
where
\begin{align}
   K&=32\pi^3\prod_{i=1}^4
   \left(\int\frac{d^3p_i}{(2\pi)^3}\,\frac{e^{-p_i^2}}{p_i^2}\right)
   (2\pi)^3\delta(p_1+p_2+p_3+p_4)
   \simeq0.289\,432,
\label{eq:(3.29)}
\\
   K'&=512\pi^3\prod_{i=1}^4
   \left(\int\frac{d^3p_i}{(2\pi)^3}\,e^{-p_i^2}\right)
   (2\pi)^3\delta(p_1+p_2+p_3+p_4)
   =\frac{1}{(4\pi)^{3/2}},
\label{eq:(3.30)}
\end{align}
and
\begin{align}
   \left\langle\mathcal{O}_2(t)\right\rangle
   &=\frac{1}{8\pi^2}
   \left(\int\frac{d^3p}{(2\pi)^3}\,\frac{e^{-2p^2}}{p^2+M^2t}
   \right)^{-1}
   -\left(\frac{8}{\pi}\right)^{1/2}M^2t
   -\frac{1}{(2\pi)^{1/2}}
\label{eq:(3.31)}
\\
   &\stackrel{t\to0}{\to}Mt^{1/2}
\label{eq:(3.32)}
\\
   &\stackrel{t\to\infty}{\to}
   \left(\frac{2}{\pi}\right)^{1/2}
   -\frac{3}{(8\pi)^{1/2}}\frac{1}{M^2t}.
\label{eq:(3.33)}
\end{align}
The asymptotic behaviors~\eqref{eq:(3.27)} and~\eqref{eq:(3.32)} show that the
initial condition of the flow is given by the parameters $\lambda_0$
and~$M$.\footnote{If one sets $M\to0$ first, Eq.~\eqref{eq:(3.26)} yields
$\langle\mathcal{O}_1(t)\rangle\stackrel{t\to0}{\to}K\lambda_0t^{1/2}$,
$\langle\mathcal{O}_1(t)\rangle\stackrel{t\to\infty}{\to}1.425\,96$, while
from~Eq.~\eqref{eq:(3.31)}, $\langle\mathcal{O}_2(t)\rangle\equiv0$.}
In~Figs.~\ref{fig:1} and~\ref{fig:2}, we depict the RG flow lines in the
space of $\langle\mathcal{O}_1(t)\rangle$ and~$\langle\mathcal{O}_2(t)\rangle$
obtained numerically. We confirmed that the point indicated by the red point
[$(\langle\mathcal{O}_1(t)\rangle,\langle\mathcal{O}_2(t)\rangle)=%
(1.425\,96,0)$] is an infrared fixed point that can be identified with the
Wilson--Fischer fixed point
($\lambda_\star/\mu=96/\sqrt{3}$ in~Eq.~\eqref{eq:(3.14)}). From the figures,
we see that $\langle\mathcal{O}_2(t)\rangle$ basically corresponds to the
relevant coupling around the fixed point; $\langle\mathcal{O}_1(t)\rangle$ to
the irrelevant coupling.

\section*{Acknowledgements}
We would like to thank Robert Harlander for helpful remarks.
The work of H. S. is supported in part by JSPS Grant-in-Aid for Scientific Research Grant Number JP16H03982.


\begin{thebibliography}{00}

\bibitem{Narayanan:2006rf} 
  R.~Narayanan and H.~Neuberger,
  JHEP {\bf 0603}, 064 (2006)
  doi:10.1088/1126-6708/2006/03/064
  [hep-th/0601210].

\bibitem{Luscher:2009eq} 
  M.~L\"uscher,
  Commun.\ Math.\ Phys.\  {\bf 293}, 899 (2010)
  doi:10.1007/s00220-009-0953-7
  [arXiv:0907.5491 [hep-lat]].

\bibitem{Luscher:2010iy} 
  M.~L\"uscher,
  JHEP {\bf 1008}, 071 (2010)
  Erratum: [JHEP {\bf 1403}, 092 (2014)]
  doi:10.1007/JHEP08(2010)071, 10.1007/JHEP03(2014)092
  [arXiv:1006.4518 [hep-lat]].

\bibitem{Luscher:2011bx} 
  M.~L\"uscher and P.~Weisz,
  JHEP {\bf 1102}, 051 (2011)
  doi:10.1007/JHEP02(2011)051
  [arXiv:1101.0963 [hep-th]].

\bibitem{Luscher:2013cpa} 
  M.~L\"uscher,
  JHEP {\bf 1304}, 123 (2013)
  doi:10.1007/JHEP04(2013)123
  [arXiv:1302.5246 [hep-lat]].

\bibitem{Wilson:1973jj} 
  K.~G.~Wilson and J.~B.~Kogut,
  Phys.\ Rept.\  {\bf 12}, 75 (1974).
  doi:10.1016/0370-1573(74)90023-4

\bibitem{Luscher:2013vga} 
  M.~L\"uscher,
  PoS LATTICE {\bf 2013}, 016 (2014)
  [arXiv:1308.5598 [hep-lat]].

\bibitem{Kagimura:2015via} 
  A.~Kagimura, A.~Tomiya and R.~Yamamura,
  arXiv:1508.04986 [hep-lat].

\bibitem{Yamamura:2015kva} 
  R.~Yamamura,
  PTEP {\bf 2016}, no. 7, 073B10 (2016)
  doi:10.1093/ptep/ptw097
  [arXiv:1510.08208 [hep-lat]].

\bibitem{Aoki:2016ohw} 
  S.~Aoki, J.~Balog, T.~Onogi and P.~Weisz,
  PTEP {\bf 2016}, no. 8, 083B04 (2016)
  doi:10.1093/ptep/ptw106
  [arXiv:1605.02413 [hep-th]].

\bibitem{Aoki:2015dla} 
  S.~Aoki, K.~Kikuchi and T.~Onogi,
  PTEP {\bf 2015}, no. 10, 101B01 (2015)
  doi:10.1093/ptep/ptv131
  [arXiv:1505.00131 [hep-th]].

\bibitem{Aoki:2016env} 
  S.~Aoki, J.~Balog, T.~Onogi and P.~Weisz,
  PTEP {\bf 2017}, no. 4, 043B01 (2017)
  doi:10.1093/ptep/ptx025
  [arXiv:1701.00046 [hep-th]].

\bibitem{Aoki:2017bru} 
  S.~Aoki and S.~Yokoyama,
  arXiv:1707.03982 [hep-th].

\bibitem{Aoki:2017uce} 
  S.~Aoki and S.~Yokoyama,
  arXiv:1709.07281 [hep-th].

\bibitem{Makino:2014taa} 
  H.~Makino and H.~Suzuki,
  PTEP {\bf 2014}, 063B02 (2014)
  Erratum: [PTEP {\bf 2015}, 079202 (2015)]
  doi:10.1093/ptep/ptu070, 10.1093/ptep/ptv095
  [arXiv:1403.4772 [hep-lat]].

\bibitem{Caswell:1974gg} 
  W.~E.~Caswell,
  Phys.\ Rev.\ Lett.\  {\bf 33}, 244 (1974).
  doi:10.1103/PhysRevLett.33.244

\bibitem{Banks:1981nn} 
  T.~Banks and A.~Zaks,
  Nucl.\ Phys.\ B {\bf 196}, 189 (1982).
  doi:10.1016/0550-3213(82)90035-9

\bibitem{Harlander:2016vzb} 
  R.~V.~Harlander and T.~Neumann,
  JHEP {\bf 1606}, 161 (2016)
  doi:10.1007/JHEP06(2016)161
  [arXiv:1606.03756 [hep-ph]].

\bibitem{Wilson:1971dc} 
  K.~G.~Wilson and M.~E.~Fisher,
  Phys.\ Rev.\ Lett.\  {\bf 28}, 240 (1972).
  doi:10.1103/PhysRevLett.28.240

\bibitem{Capponi:2015ucc} 
  F.~Capponi, A.~Rago, L.~Del Debbio, S.~Ehret and R.~Pellegrini,
  PoS LATTICE {\bf 2015}, 306 (2016)
  [arXiv:1512.02851 [hep-lat]].

\end{thebibliography}
\end{document}